\documentclass[a4paper]{article}
\usepackage{lineno,hyperref}
\usepackage[utf8]{inputenc}
\usepackage{rotating}
\usepackage{pdflscape}
\usepackage{placeins}
\usepackage{graphicx}
\usepackage{subcaption} 
\usepackage{subfiles}
\usepackage{tikz}
\usetikzlibrary{positioning, fit, arrows.meta, shapes}

\usepackage{pgfplots}
\pgfplotsset{compat=1.9}
\usepackage{import}
\usepackage{adjustbox}
\usepackage{textcomp}
\usepackage{blindtext}
\usepackage{authblk}

\usepackage{color, colortbl}
\definecolor{Gray}{gray}{0.9}
\definecolor{LightCyan}{rgb}{0.88,1,1}

\usepackage{float}
\usepackage{longtable}
\usepackage{multirow}

\usepackage{amsmath}
\usepackage{amssymb}
\usepackage{amsthm}
\usepackage{nameref}
\usepackage{cleveref}
\usepackage{booktabs}	

\usepackage{url}
\usepackage{epsfig}
\usepackage{placeins} 
\usepackage{longtable} 
\usepackage{marginnote} 
\usepackage{geometry}
\usepackage[round]{natbib}
\setcitestyle{authoryear,open={(},close={)}}

\title{Automated Identification of Vulnerable Devices in Networks using Traffic Data and Deep Learning}

\author[1]{Jakob Greis *}
\author[1]{Artem Yushchenko}
\author[1]{Daniel Vogel}
\author[1,2]{Michael Meier}
\author[1]{Volker Steinhage \footnote{Corresponding authors: \{greis,steinhage\}@cs.uni-bonn.de}}
\affil[1]{University of Bonn, Institute of Computer Science IV \\Friedrich-Hirzebruch-Allee 8, Bonn 53115, Germany}
\affil[2]{Fraunhofer Institute for Communication, Information Processing and Ergonomics (FKIE) Friedrich-Hirzebruch-Allee 8, Bonn 53115, Germany}

\begin{document}

\maketitle
\begin{quotation}
\section*{\centering \begin{normalsize}Abstract\end{normalsize}}
Many IoT devices are vulnerable to attacks due to flawed security designs and lacking mechanisms for firmware updates or patches to eliminate the security vulnerabilities.
Device-type identification combined with data from vulnerability databases can pinpoint vulnerable IoT devices in a network and can be used to constrain the communications of vulnerable devices for preventing damage. 
In this contribution, we present and evaluate two deep learning approaches to the reliable IoT device-type identification, namely a recurrent and a convolutional network architecture. Both deep learning approaches show accuracies of 97\% and 98\%, respectively, and thereby outperform an up-to-date IoT device-type identification approach using hand-crafted fingerprint features obtaining an accuracy of 82\%. The runtime performance for the IoT identification of both deep learning approaches outperforms the hand-crafted approach by three magnitudes. Finally, importance metrics explain the results of both deep learning approaches in terms of the utilization of the analyzed traffic data flow. 

\end{quotation}
\clearpage

\section{Introduction}
\label{sec:introduction}
The Internet of Things (IoT) is a collective term for technologies that enable physical and virtual objects to be networked in order to make them work together. 
An IoT device is a single agent in an IoT network, such as those devices found in a smart home network, e.g., smart doorbells, smart locks, smart heating control, smart refrigerators, smart speakers and others. 
The total number of IoT devices installed worldwide is estimated at 75.44 billion by 2025, \citep{statista}. 


IoT devices may suffer from poor security designs and missing firmware updates or patches to eliminate security vulnerabilities \citep{Nakajima}. Therefore, IoT devices with unpatched vulnerabilities may coexist with other devices in home and office networks thereby opening up new attack vectors. 

Securing such networks showing the presence of vulnerable devices can be implemented in three steps: (1) identification of the IoT device types, (2) vulnerability assessment based
on querying repositories like the CVE database \citep{CVE} for vulnerability reports, and (3) application of a mitigation strategy to maintain as much functionality as possible while minimizing the risk of harm \citep{sentinel2017}.




\subsection{Contributions}

This study contributes a methodological evaluation of two deep learning approaches to implement the first step in the aforementioned processing pipeline, i.e., the automated identification IoT device types. More specifically, we apply a deep Convolution Neuronal Networks as well as a deep Recurrent Neural Network to the automated identification IoT device types based on raw traffic data. The evaluation shows that both deep learning approaches not only outperform the up-to-date approach proposed by \citet{sentinel2017} but also apply representation learning to learn automatically relevant features for the device type identification from raw data and thereby solve the problem of hand-designing such features \citep{RepresentationLearning}.



\section{Related Work}
\label{sec:goalsandcontributions}

\label{sec:iotdeviceidentification}



Approaches to the automated identification of IoT devices based on traffic data can be divided into three different categories:  

\subsection{Targeting device-specific hardware \& software configuration}
Approaches of this category focus on hardware or software specific characteristics of individual IoT devices, e.g. \citet{Gligor2008fingerprinting} and \citet{cache2006fingerprinting}). But the usage of drivers and hardware characteristics to identify IoT devices can be insufficient since the same hardware components and drivers may be used in a wide variety of different device types. However, the usage of hardware specific features like clock offset or unique network interface cards \citep{specificFingerprinting2005} is too specific if one is interested in the identification the device types since vulnerability does not affect a single device but always the entire device type. 
This issue also affects approaches that use fingerprints based on wireless communication since IoT devices may use wireless communication but also may use LAN or proprietary communication techniques.

\subsection{Targeting Network Communications Fingerprints }
\label{sec:sentinel}
This second category includes all methods where additional communications fingerprints are derived from traffic data. \citet{sentinel2017}  derived manually characteristic features while \citep{audi2019} employed broadcast frequencies and unsupervised learning. Both approaches refer mainly to the (not encrypted) communication characteristics and less to the content of the respective (encrypted) packages. But this way, not all available information is used. Even encrypted data contains worth-full patterns for the identification of ioT devise types \citep{WeiWang2017}.


\subsection{Targeting Raw Network Data} 
\label{sec:weiwang}
Approaches of this category work on raw and unprocessed data. \citet{WeiWang2017} classify malware just using raw traffic data without deriving malware characteristic features manually. Instead, they use raw traffic data packets to train a deep convolutional neural network to identify different classes of malware. In order to improve the accuracy of the network, they suggest to choose sessions instead of flows or complete recordings for training and predictions of the neural network.

\section{Material and Methods}
\label{sec:methodology}

The approach of the IoT SENTINEL project \citep{sentinel2017} to secure IoT networks with vulnerable devices and its three-step processing pipeline (IoT device type identification, vulnerability assessment, mitigation strategy) forms the appropriate framework for our study on improvement of the first step, i.e., the IoT device type identification. Furthermore, we will use the identification results of the IoT Sentinel project as a baseline for evaluating our proposed deep learning approaches. Therefore, we will also use the data set of the IoT SENTINEL project.

\subsection{SENTINEL Data set}
\label{sec:dataset}

The device-type identification of the IoT SENTINEL approach is based on monitoring the communication during the setup processes of the devices to generate device-specific
fingerprints. For each device, the setup process was performed 20 times. After each setup, a hard reset of the device was performed to allow reuse. The IoT SENTINEL data set was recorded for 27 IoT device types that are summarized in table \ref{tab:SentinelDevices}. 
\\

The PCAP recordings of the IoT SENTINEL data set were collected using a Linux laptop as a gateway. Via this gateway, a man in the middle attack collected all ongoing packets while the setup procedures. The data set covered a wide range of common smart home devices, covering the following areas: intelligent lighting, home automation, security cameras, household equipment and condition monitoring devices. Most of the tested devices were connected to the user's network via WiFi or Ethernet. However, some devices used other IoT protocols such as ZigBee or Z-Wave for indirect connection to the network via an Ethernet or WiFi hub device. 

\subsubsection{IoT SENTINEL Features}

For each new device $n$ packets $p_1, p_2, ... p_n$ are recorded during its setup phase. \citet{sentinel2017} derive 23 manually designed packet features that works independently of packed payload data and ensures that the features can also be extracted for encrypted communication. An overview over all features are given in table \ref{tab:SentinelFeatures}. 

Therefore, a device setup yields for each recorded packet $p_i$ a vector $(f_{1,i}, f_{2,i}, ... f_{23,i})$ and for each device a fingerprint in terms of a $23 \times n$ matrix $F$ with each column representing a packet received with order $i \in \{1, 2, ...n\}$ and each row representing a packet feature.

\begin{table}[H]
\centering
  \caption{IoT device types of the IoT SENTINEL dataset \citep{sentinel2017}.}
  \label{tab:SentinelDevices}
\begin{tabular}{ll}
\toprule
Identifier & Device Model\\
\midrule
Aria & Fitbit Aria WiFi-enabled scale\\
HomeMaticPlug & Homematic pluggable switch HMIP-PS\\
Withings & Withings Wireless Scale WS-30\\
MAXGateway & MAX! Cube LAN Gateway for MAX! Home automation sensors\\
HueBridge & Philips Hue Bridge model 3241312018 \\
HueSwitch & Philips Hue Light Switch PTM 215Z\\
EdnetGateway & Ednet.living Starter kit power Gateway\\
EdnetCam & Ednet Wireless indoor IP camera Cube\\
EdimaxCam & Edimax IC-3115W Smart HD WiFi Network Camera\\
Lightify & Osram Lightify Gateway\\
WeMoInsightSwitch & WeMo Insight Switch model F7C029de\\
WeMoLink & WeMo Link Lighting Bridge model F7C031vf\\
WeMoSwitch & WeMo Switch model F7C027de\\
D-LinkHomeHub & D-Link Connected Home Hub DCH-G020\\
D-LinkDoorSensor & D-Link Door Window sensor\\
D-LinkDayCam & D-Link WiFi Day Camera DCS-930L\\
D-LinkCam & D-Link HD IP Camera DCH-935L\\
D-LinkSwitch & D-Link Smart plug DSP-W215\\
D-LinkWaterSensor & D-Link Water sensor DCH-S160\\
D-LinkSiren & D-Link Siren DCH-S220\\
D-LinkSensor & D-Link WiFi Motion sensor DCH-S150\\
TP-LinkPlugHS110 & TP-Link WiFi Smart plug HS110\\
TP-LinkPlugHS100 & TP-Link WiFi Smart plug HS100\\
EdimaxPlug1101W & Edimax SP-1101W Smart Plug Switch\\
EdimaxPlug2101W & Edimax SP-2101W Smart Plug Switch\\
SmarterCoffee & Smarter SmarterCoffee coffee machine SMC10-EU\\
iKettle2 & Smarter iKettle 2.0 water kettle SMK20-EU\\
\bottomrule
\end{tabular}
\end{table}

\begin{table}[H]
\centering
  \caption{Description of the 23 manually designed packet features of IoT SENTINEL. All features are binary except those marked as “(integer)” \citep{sentinel2017}.}
  \label{tab:SentinelFeatures}
\begin{tabular}{lcl}
\toprule
Type                            & \# Features    & Features \\
\midrule
Link layer protocol             & 2             &   ARP / LLC \\
Network layer protocol          & 4             & IP / ICMP / ICMPv6 / EAPoL \\
Transport layer protocol        & 2             & TCP / UDP \\
Application layer protocol      & 8             & HTTP / HTTPS / DHCP / BOOTP /\\
                                &               & SSDP / DNS / MDNS / NTP \\
IP options                      & 2             & Padding / RouterAlert \\
Packet content                  & 2             & Size (integer) / Raw data \\
IP address                      & 1             & Destination IP counter (integer) \\
Port class                      & 2             & Source (integer) / Destination (integer) \\
\bottomrule
\end{tabular}
\end{table}



\subsubsection{Traffic Samples of SENTINEL Data set}
\label{sec:sentinel}

Given the 27 device types and 20 setups per device type the IoT SENTINEL data set comprises 540 device setups yielding 18923 feature sets as seen in \cref{tab:SentinelVolume}. Device fingerprints have variable size (i.e. number of packages) and consecutive identical packets are discarded from the feature set perspective.

\begin{table}[H]
\centering
  \caption{Volume of the IoT SENTINEL data set \citep{sentinel2017}.}
  \label{tab:SentinelVolume}
\begin{tabular}{llll}
\toprule
Data set Name & Total samples & device types & format \\
\midrule
SENTINEL PCAP              &          540 &              27 &             PCAP \\
SENTINEL Feature         &          18923 &              27 &            CSV \\
\bottomrule
\end{tabular}
\end{table}

While IoT SENTINEL uses the manually designed fingerprints of the 540 setups for device type identification (cf. section \ref{sec:devicetypeidentification}), we will apply two deep learning approaches to the original byte streams of the setups (cf. sections \ref{sec:cnn}, \ref{sec:LSTM}).  

\subsubsection{Device Type Identification in IoT SENTINEL}
\label{sec:devicetypeidentification}
The IoT SENTINEL approach of \citet{sentinel2017} proposes a two-step approach to the IoT device type identification. First, a Random Forest classification algorithm \citep{Breiman01} is applied in a “one classifier per device type” approach. Given 27 different device types, 27 Random Forests are trained, each on the fingerprints of the devices of one device type. In the identification phase, each of these 27 trained Random Forests yield binary decision whether an unknown input fingerprint belongs to the corresponding device type or not. But an unknown fingerprint may match several device types during the “one classifier per device type” identification process yielding multiple device type candidates. Therefore, the similarity of the unknown fingerprint with the fingerprints from each device type candidate is used to make the final decision. The fingerprint comparison is done by computing the Damerau-Levenshtein edit distance \citep{damerau} that measures the numbers of insertions, deletions, substitutions and immediate transpositions of characters.


\subsection{CNN-based Device Type Identification}
\label{sec:cnn}

Convolution Neural Networks \citep{LeCun}, or CNNs for short, form currently one of the most popular approaches to representation learning, especially in the fields computer vision and visual recognition.
Therefore, the application of a CNN architecture to the raw network traffic data of the IoT SENTINEL data set will result in an end-to-end learning not only of the identification of the IoT device types but also of the features relevant for this identification process. 
Such an approach, i.e., network traffic analysis using CNNs has been first proposed by \citet{WeiWang2017} for malware traffic detection and classification.  

\begin{figure}[H]
\begin{center}
  \begin{subfigure}{0.24\columnwidth}
  \centering
  \includegraphics[width=0.9\linewidth]{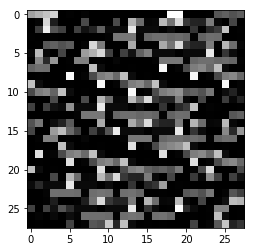}
  \end{subfigure}
  \begin{subfigure}{0.24\columnwidth}
  \centering
  \includegraphics[width=0.9\linewidth]{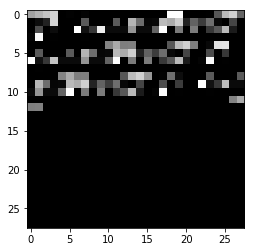}
  \end{subfigure}
  \begin{subfigure}{0.24\columnwidth}
  \centering
  \includegraphics[width=0.9\linewidth]{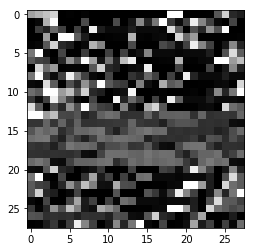}
  \end{subfigure}
  \begin{subfigure}{0.24\columnwidth}
  \centering
  \includegraphics[width=0.9\linewidth]{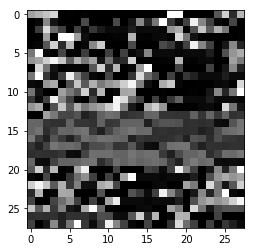}
  \end{subfigure}
\end{center}
\caption{Gray-value images generated from the first 784 bytes of the device setups of EdnetGateway, HueSwitch, MAXGateway and SmarterCoffee (f.l.t.r.).}
\label{fig:pictures}
\end{figure}

Since most CNN architectures are designed to process images, the first step is a simple transformation of the raw traffic data into images. 
Following the approach of \citet{WeiWang2017} and assuming that a significant long part of the setup traffic data will suffice for the device type identification, we transform the raw traffic data of each device setup the first 784 bytes into a gray-value image of size 28 $\times$ 28 pixels where each pixel depicts a gray-value between 0 (black) and 255 (white). 
If a setup shows less than 784 bytes, the values of the remaining pixels are set to 0 (black) (cf. figure \ref{fig:pictures}).

The size of the processed data and the size of the generated images are both comparable to those in the classic MNIST data set of handwritten digits \citep{MNIST}. Therefore, \cite{WeiWang2017} proposed a CNN architecture similar to that of LeNet-5 \citep{LeCun95}. 
We also designed a CNN architecture similar to that of LeNet-5 but showing more channels.  

The CNN first reads traffic images of size of 28 $\times$ 28  $\times$ 1. 
The first convolution layer C1 performs a convolution operation with 32 kernels of size of 5 $\times$ 5 without padding. 
The results of the convolution layer C1 are 32 feature maps of size of 24 $\times$ 24. 
These feature maps are down-sampled via a 2 $\times$ 2 average pooling operation resulting in 32 feature maps of size of 12 $\times$ 12. 
The kernel size of the second convolution layer C2 is also 5 $\times$ 5 without padding but has 64 channels. 
This results in 64 feature maps of size of 8 $\times$ 8 and after a second average pooling in 64 feature maps of size of 4 $\times$ 4. 
The flattened version of these 1024 neurons is used as input for two fully connected layers with 120 and 84 neurons resulting in a Softmax output layer of 27 neurons for the 27 device types (cf. table \ref{tab:CNN_struck}).

\begin{table}[H]
\center
\caption{The architecture of the CNN-based IoT device type identification.}
\label{tab:CNN_struck}
\begin{tabular}{llll}
\toprule
Input    & Layer       &             &               \\
\hline
28*28*1  & Conv        & 32 filter   & 5 filter size \\
24*24*32 & Avg Pooling & 2*2 pooling &               \\
12*12*3  & Conv        & 32 filter   & 5 filter size \\
8*8*64   & Avg Pooling & 2*2 pooling &               \\
4*4*64   & Flatten     &             &               \\
1024     & Dense       & 120 neurons &               \\
120      & Dense       & 84neurons   &               \\
84       & Softmax     & 27 neurons  &      \\
\bottomrule          
\end{tabular}
\end{table}


\subsection{LSTM-based Device Type Identification}
\label{sec:LSTM}

Network traffic data is a sequential flow of data. 
First of all, the IP header files contain key information about the subsequent data transmission. 
Then, for many packets not only the content is decisive but also their local and global context within a session. 
This is strongly reminiscent to Natural Language Processing (NLP), since natural language is spoken and written as sequences of words and sentences.

Long short-term memory networks, or LSTMs for short, are a variation of artificial recurrent neural networks (RNN) used for processing sequential data as in NLP \citep{hochreiter1997}. 
In contrast to other RNN architectures, LSTMs can efficiently keep track of arbitrary long-term dependencies in the input sequences. 
Additionally, the LSTM approach overcomes the so-called vanishing gradient problem that can be encountered when training traditional RNNs with Backpropagation Through Time (BPTT) (e.g., \citet{BPTT}. 

But standard BPTT requires saving the entire history of inputs and internal activations of the forward pass for use in the backpropagation step. 
This requirement is computationally expensive and memory intensive for long input sequences. 
Therefore, \citet{Williams95gradient-basedlearning} and \citet{pmlr-v28-sutskever13} proposed Truncated Backpropagation Through Time (TBPTT) as an efficient approximation where the input sequence is treated as fixed length subsequences, so-called chunks. 

For an LSTM-based device type identification using the raw network traffic data, the LSTM architecture shows first two LSTM layers followed by a 20\% dropout to prevent overfitting and than two fully connected dense layers classifying the results (cf. table \ref{tab:lstm_struck}). 
To provide a comparative evaluation of the CNN-based and the LSTM-based device type identification, the input of the LSTM architecture is also restricted to the first 784 bytes of the setup traffic.

\begin{table}[H]
\center
\caption{The architecture of the LSTM-based IoT device type identification.}
\label{tab:lstm_struck}
\begin{tabular}{llll}
\toprule
Input & Layer   &            &  \\
784   & LSTM    & 128 units  & returning sequence \\
128   & LSTM    & 128 units  &  \\
128   & DropOut & 0.2        &  \\
128   & Dense   & 64 neurons &  \\
64    & Softmax & 27 neurons & \\
\bottomrule
\end{tabular}
\end{table}

\subsection{Feature Importance}

While deep neural networks in general have turned out to be extremely powerful in the development of intelligent systems they suffer from the lack of transparency due to their multilayer nonlinear structure, limiting the interpretability of the solution. 
In response, various methods have been proposed to explain the results of such complex models, e.g., Mean Decrease Accuracy (MDA) \citep{Breiman01}, Conditional Permutation Importance (CPI) \citep{strobl} or SHapley Additive exPlanations (SHAP) \citep{lundberg2017shap}. 
We employ the SHAP approach, since SHAP measures feature importance at a local level to understand each particular decision and at a global level to understand the complete model.


\section{Evaluation and Discussion}
\label{sec_evaluationanddiscussion}


All code of the implementation and evaluation was written in Python 3. 
For the implementation of the machine learning methods we used the Python library Scikit. 
Furthermore we used the Python framework Keras for the implementation of the deep learning models. 
All experiments were performed on Colab. 
Google Colaboratory is a research project for prototyping machine learning models on powerful hardware. 
It provides a Jupyter notebook environment for interactive development that runs in the cloud. 
The instance we used in our experiment had a NVIDIA Tesla T80 GPU, 16GB Ram, and an Intel(R) Xeon processor.

\subsection{Network Traffic Representation}

To apply deep learning to the network traffic data, first some initial data preprocessing has to be employed. 
Second, the continuous traffic stream must be split into discrete units of a certain level of granularity. 

Our preprocessing follows the propsal of \citet{WeiWang2017}, i.e., IP and MAC information is removed from the traffic data by randomization, usually called traffic anonymization or sanitization \citep{Koukis} to avoid negative impact of this information to the feature learning of the neural networks. 
Additionally, packets that have no application layer as well as empty and duplicated files are removed.

With respect to the granularity level of the traffic representation, there are several ways to partition the traffic stream including TCP connection, flow, session, service/protocol, or used host \citep{dainotti2012issues}. 
In this study, sessions are the granularity level of choice since sessions provide for training of deep learning approaches on the one hand a sufficient high number of samples and on the other hand sufficient information per sample. 
While flow is only a uni-directional transfer between a pair of (IP address, port) using the same transport-level protocol, type of service, and input interface, a session is the bi-directional transfer between this pair of (IP Address, Port) and therefor the sum of the flows in both directions. 
Given 20 setups per device type and therefore 540 setups for all 27 device types, 15023 sessions are now available for training and testing of both deep learning approaches.



\subsubsection{Class Weighting}
\label{sec:classweighting}
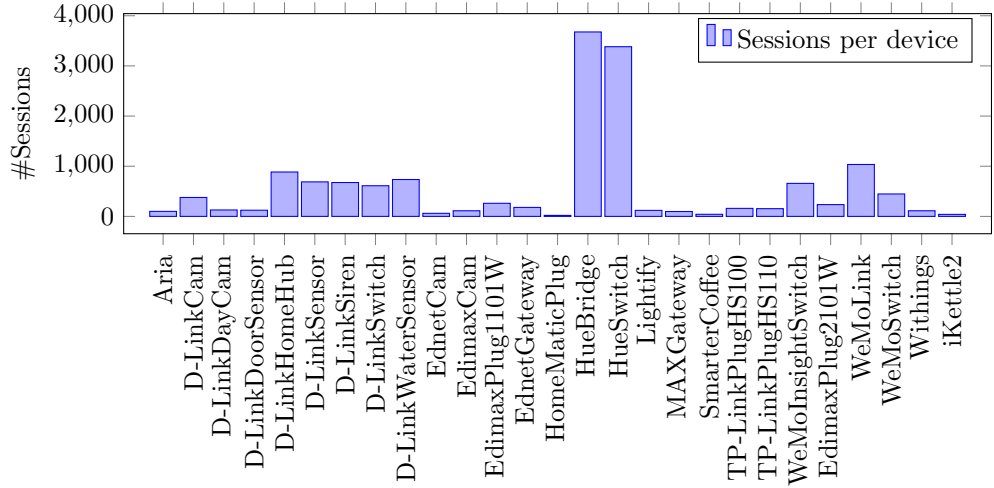
\begin{figure}
\begin{tikzpicture}
\begin{axis}[
    width=13cm,
    height=4.5cm,
    enlarge x limits  = 0.05,
    ybar,
    legend style={at={(0.82,0.98)},
      anchor=north,legend columns=-1},
    ylabel={\#Sessions},
xticklabel style={rotate=90},
    symbolic x coords={ Aria, D-LinkCam, D-LinkDayCam, D-LinkDoorSensor, D-LinkHomeHub,  D-LinkSensor, D-LinkSiren,D-LinkSwitch, D-LinkWaterSensor,   EdnetCam,   EdimaxCam,  EdimaxPlug1101W, EdnetGateway,  HomeMaticPlug, HueBridge,   HueSwitch,  Lightify,   MAXGateway,  SmarterCoffee, TP-LinkPlugHS100, TP-LinkPlugHS110, WeMoInsightSwitch, EdimaxPlug2101W, WeMoLink, WeMoSwitch, Withings, iKettle2},
    xtick=data,
    ]
\addplot coordinates {(Aria,100) (D-LinkCam,377) (D-LinkDayCam,128)(D-LinkDoorSensor,123) (D-LinkHomeHub, 885) (D-LinkSensor,688) (D-LinkSiren,675) (D-LinkSwitch,611)(D-LinkWaterSensor,735)(EdnetCam,61)(EdimaxCam,112)(EdimaxPlug1101W,261)(EdnetGateway,179)(HomeMaticPlug, 20)(HueBridge, 3677)(HueSwitch, 3383)(Lightify,120)(MAXGateway,97)(SmarterCoffee,42)(TP-LinkPlugHS100, 160)(TP-LinkPlugHS110,154)(WeMoInsightSwitch,658) (EdimaxPlug2101W, 235)(WeMoLink, 1035)(WeMoSwitch, 447)(Withings,112)(iKettle2,40)};
\legend{Sessions per device}
\end{axis}
\end{tikzpicture}
\caption{The number of sessions of each device type.}
\label{fig:SessionStats1}
\end{figure}

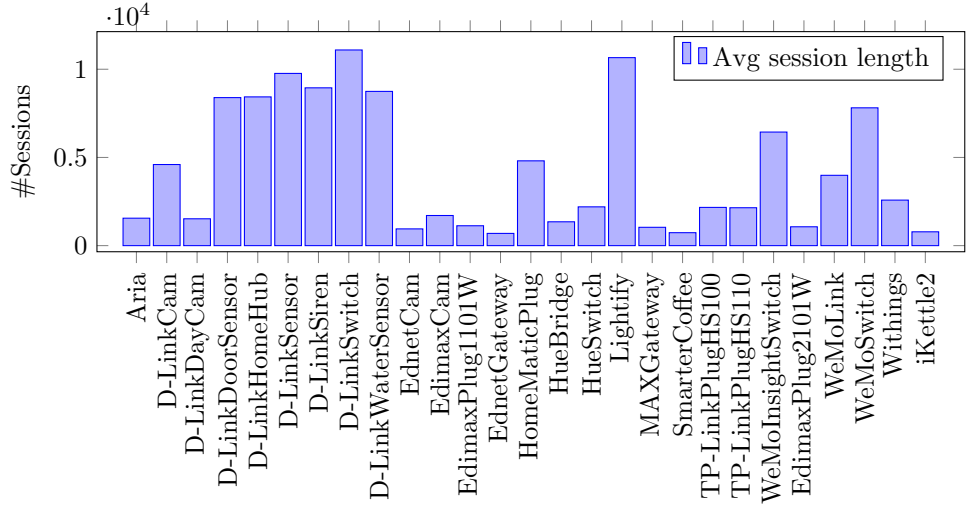
\begin{figure}[H]
\begin{tikzpicture}
\begin{axis}[
    width=13cm,
    height=4.5cm,
    enlarge x limits  = 0.05,
    ybar,
    legend style={at={(0.82,0.98)},
      anchor=north,legend columns=-1},
ylabel={\#Sessions},
xticklabel style={rotate=90},
    symbolic x coords={ Aria, D-LinkCam, D-LinkDayCam, D-LinkDoorSensor, D-LinkHomeHub, D-LinkSensor, D-LinkSiren, D-LinkSwitch, D-LinkWaterSensor,   EdnetCam,   EdimaxCam,  EdimaxPlug1101W, EdnetGateway,  HomeMaticPlug, HueBridge,   HueSwitch,  Lightify,   MAXGateway,  SmarterCoffee,TP-LinkPlugHS100,  TP-LinkPlugHS110,  WeMoInsightSwitch, EdimaxPlug2101W, WeMoLink, WeMoSwitch, Withings, iKettle2},
    xtick=data,
    ]
\addplot coordinates {
(Aria,1556.14)
(D-LinkCam,4598.1)
(D-LinkDayCam,1521.81)
(D-LinkDoorSensor,8391.6) (D-LinkHomeHub, 8432.44) (D-LinkSensor,9763.11) (D-LinkSiren,8945.92) (D-LinkSwitch,11091.9)(D-LinkWaterSensor,8743.91)(EdnetCam,949.23)(EdimaxCam,1708.23)(EdimaxPlug1101W,1126.7)(EdnetGateway,693.721)(HomeMaticPlug, 4807)(HueBridge, 1351.44)(HueSwitch, 2198.95)(Lightify,10657.4)(MAXGateway,1039.49)(SmarterCoffee,734.095)(TP-LinkPlugHS100, 2167.66)(TP-LinkPlugHS110,2148.52)(WeMoInsightSwitch,6438.62) (EdimaxPlug2101W, 1068.91)(WeMoLink, 3986.79)(WeMoSwitch, 7812.38)(Withings,2580.81)(iKettle2,784)};
\legend{Avg session length}
\end{axis}
\end{tikzpicture}
\caption{The average session length of each device type.}
\label{fig:SessionStats2}
\end{figure}

The data set shows a highly unbalanced data distribution with respect to the IoT device types (cf. figures \ref{fig:SessionStats1} and \ref{fig:SessionStats2}). 
The average number of sessions per device type is 559.8 with a standard deviation of 904.64. 
Additionally, 20\% of all sessions can be found for the two device types \textit {HueBridge} and \textit{HueSwitch}. 
Training with unbalanced data could yield lower identification accuracy for device types that are underrepresented in the training data. 
Furthermore, device types that are overrepresented in the training can be overfitted in training.
Therefore, we apply class weighting by assigning individual weights to all device types according to their number of sessions in the training data. 
For the weight calculation we used the following definition:

\[ w_d = \dfrac{max_N}{n_d}, \]

where $w_d$ is the weight of device type $d$, $max_N$ is the maximum number of sessions found for a device type, and $n_d$ is the number of sessions of device type $d$.

\subsection{Comparative Evaluation of CNN vs LSTMs}

All samples are split into train and test data sets with a train-test ratio of 0.8 : 0.2 using random selection to ensure that the train and test data sets are representative of the original data set. 
Table \ref{tab:SessionSetUpAutomaticFeatures} summarizes the identification performances of the CNN model and the LSTM model applied to complete setups and to single sessions using accuracy, precision and recall values of the average top-1 scores with
\[
accuracy = \dfrac{TP + TN}{ P + N},
\hspace{12pt}
recall = \dfrac{TP}{TP + FN},
\hspace{12pt}
precision = \dfrac{TP}{ TP + FP}
\]
for identification results $P$ = all positives, $N$ = all negatives, $TP$ = true positives, $TN$ = true negatives, $FP$ = false positives, $FN$ = false negatives.

The employed CNN shows the LeNet-5 architecture adapted as described in section \ref{sec:cnn}. 
The employed LSTM uses TBPTT as described in section \ref{sec:LSTM}. 
Table \ref{tab:SessionSetUpAutomaticFeatures} shows that both approaches show by far better identification results when applied to single sessions instead of complete setups.

\begin{table}[H]
\center
\caption{IoT device identification performance of the CNN model and the LSTM model applied to complete setups and to sessions.}
\begin{tabular}{lllr}
\toprule
Model / Granularity & Accuracy & Precision & \multicolumn{1}{l}{Recall} \\ 
\hline 
\multicolumn{1}{l|}{CNN on setups}     & 0.684   & 0.606     & 0.549                     \\
\multicolumn{1}{l|}{LSTM on setups}   & 0.633   & 0.639     & 0.652                      \\

\multicolumn{1}{l|}{CNN on session}      &\cellcolor{Gray}0.982    &\cellcolor{Gray}0.985  &\cellcolor{Gray}0.983           	\\
\multicolumn{1}{l|}{LSTM on session}      & 0.971    & 0.963     & 0.972                  	\\
         
\bottomrule     
\end{tabular}
\label{tab:SessionSetUpAutomaticFeatures}
\end{table}

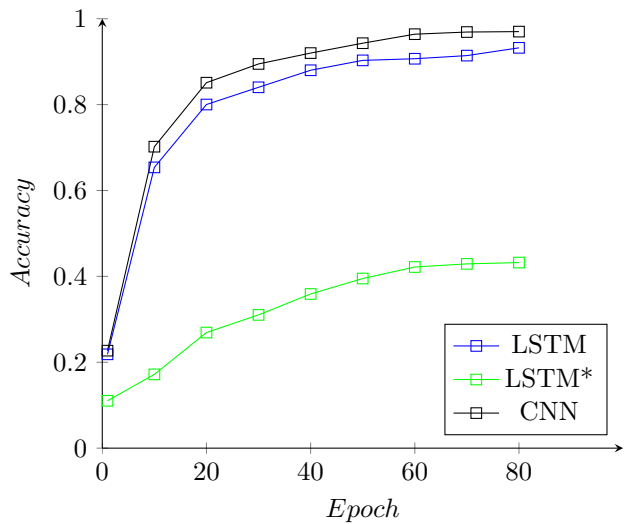
\begin{figure}[H]
\center
	\begin{tikzpicture}
		\begin{axis}[
		    axis lines = left,
		    legend pos=south east,
		    xmin=0, xmax=100,
		    ymin=0, ymax=1,
		    xtick={0,20,40,60,80},
		    ytick={0,0.2,0.4,0.6,0.8,1},
		    xlabel = $Epoch$,
		    ylabel = {$Accuracy$},
		]
		\addplot[
		    color=blue,
		    mark=square,
		 ]
		 coordinates {
		    (1,0.2185)(10,0.6538)(20,0.8002)(30,0.8403)(40,0.8800)(50,0.9030)(60,0.9068)(70,0.9141)(80,0.9322)
		 };

	\addplot[
		    color=green,
		    mark=square,
		 ]
		 coordinates {
		    (1,0.1103)(10,0.1718)(20,0.2692)(30,0.3103)(40,0.3590)(50,0.3950)(60,0.4218)(70,0.4291)(80,0.4322)
		 };
	
		\addplot[
		    color=black,
		    mark=square,
		 ]
		 coordinates {
		    (1,0.2270)(10,0.7020 )(20,0.8510)(30,0.8947)(40,0.92)(50,0.943)(60,0.964)(70,0.969)(80,0.970)
		 };
	
		 \legend{LSTM,LSTM*,CNN}	
		\end{axis}
	\end{tikzpicture}
       \caption{The accuracy vs. epoch plots of the LeNet-5 CNN, the LSTM with TBTT and the LSTM* without TBTT.}
       \label{fig:epochAccRawData}
\end{figure}

The accuracy vs. epoch plot of figure \ref{fig:epochAccRawData} depicts additional aspects of the deep learning-based device identification based on session classification. 
The modified LeNet-5 trained on 80 epochs achieves the best accuracy with 98\%. 
Already after 25 epochs it reaches an accuracy of 90\%. 
The LSTM benefits significantly from more epochs. 
This dependency is due to the different number of parameters to be trained. 
While the LeNet-5 only has to learn a few hundred thousand parameters, the LSTM has several million parameters to be learned. 

Figure \ref{fig:epochAccRawData} compares also the LSTM approach with and without TBPTT showing the significant added value of TBPTT.

\begin{figure}[H]
     \vspace{-100pt}
     \hspace{-70pt}
     \begin{subfigure}{270pt}
		\adjustbox{height=350pt,width=350pt}{\subimport{images/plt/}{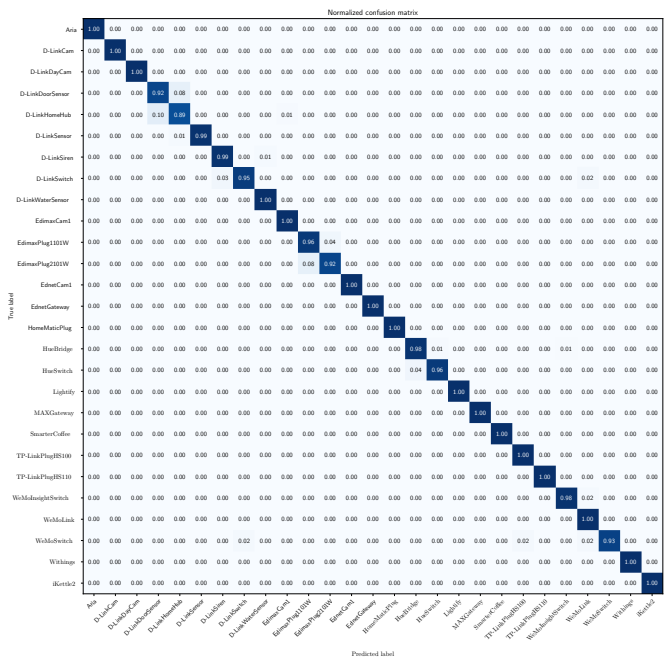}}
         \label{fig:Confusion_Matrix_CNN}
     \end{subfigure}
     \begin{subfigure}{270pt}
		\adjustbox{height=350pt,width=350pt}{\subimport{images/plt/}{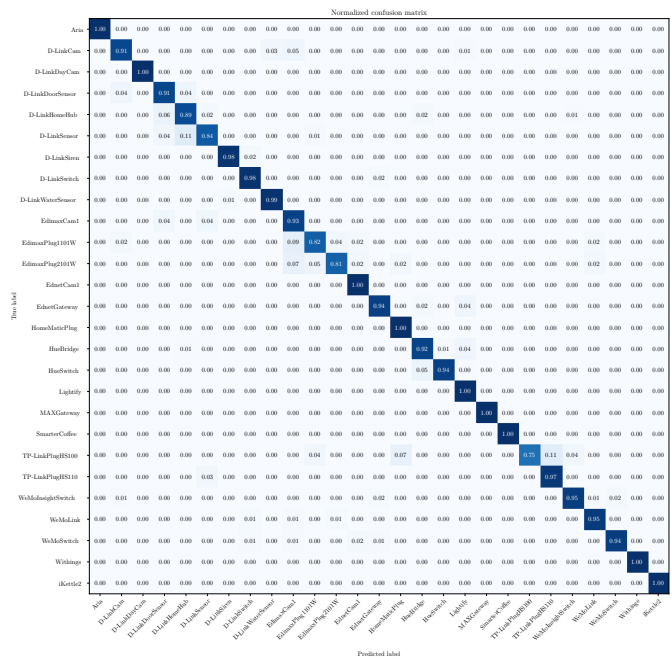}}
         \label{fig:Confusion_Matrix_LSTM}
     \end{subfigure}
     \vspace{-50pt}
  \caption{Confusion matrices of IoT device type identification using CNN(l.) and LSTM(r.), respectively.}
  \label{fig:ConfusionMatrix}
\end{figure}

The normalized confusion matrices of figure \ref{fig:ConfusionMatrix} and the cutouts of figure \ref{fig:s_ConfMatrix} show that some device groups (e.g., DLink, TP-Link) are particularly difficult to distinguish for the LSTM: 
Devices of the same manufacturer are more difficult to distinguish because they share indistinguishable software and exchange data with similar servers and therefore have comparable communication patterns. 
In the CNN's confusion matrix we see that these devices can be distinguished better indicating that some device type specific patterns are session assignable.

\begin{figure}[H]
    \begin{center}
     \begin{subfigure}{\textwidth}
	\centering
          \begin{tabular}{ c|c|c|c|c|}
	    D - Link   & DoorSensor& HomeHub & Sensor & Siren \\ \hline
	    DoorSensor      & 0.92  & 0.08 &   0.00 & 0.00   \\ \hline
	    HomeHub       & 0.10  & 0.89 &   0.00 & 0.00 \\ \hline
	    Sensor      & 0.00  & 0.01 &   0.99 & 0.00    \\ \hline
	    Siren       & 0.00  &0.00 &	0.00 &	0.99    \\   \hline
	    \end{tabular}
         \caption{CNN}
     \end{subfigure}
     \begin{subfigure}{\textwidth}
	\centering
          \begin{tabular}{ c|c|c|c|c|}
	    D - Link   & DoorSensor& HomeHub & Sensor & Siren \\ \hline
	    DoorSensor      & 0.91  & 0.04 &   0.00 & 0.00   \\ \hline
	    HomeHub       & 0.06  & 0.89 &   0.02 & 0.00 \\ \hline
	    Sensor      & 0.04  & 0.11 &   0.84 & 0.00    \\ \hline
	    Siren       & 0.00  &0.00 &	0.00 &	0.98    \\   \hline
	    \end{tabular}
         \caption{LSTM}
     \end{subfigure}
    \caption{Cutouts of the confusion matrices of fig. \ref{fig:ConfusionMatrix}: CNN avg: 0.9475, LSTM avg: 0.9050}
    \label{fig:s_ConfMatrix}
    \end{center}
\end{figure}

\subsubsection{Runtime Performance}
\label{sec:performance}
Figure \ref{fig:runTime} reports the average runtime for IoT device type identification for both deep learning approaches in comparison with the original Sentinel approach based on random forest classification and edit distance discrimination. 
Both deep learning approaches show better runtime performances of about three magnitudes.

\begin{figure}[H]
	\centering
     \begin{tabular}{ c|c|c}
	    Methode   & Mean& StDev  \\ \hline
	    SENTINEL      & 150 ms  & 170.6 ms    \\ \hline
	    CNN      & 0.2 ms  & 0.09 ms     \\ \hline
	    LSTM       & 0.67 ms  &0.14 ms    \\   \hline
	\end{tabular}
	\label{fig:PerformanceClassification}
	\caption{The average time for device-type identification per method.}
    \label{fig:runTime}
\end{figure}

\subsection{Feature Importance Analysis}
\label{sec:featureanalysis}

In order to gain insight into the decision making process of both deep learning approaches, the feature importance method SHAP \cite{lundberg2017shap}, specifically the GradientExplainer of the \textit{shap} Python package, is utilized. 
It computes the importance of features, and in our case bytes in the network traffic, by averaging their contribution to the prediction across the entire Sentinel data set.

As described in section \ref{sec:cnn}, in the CNN-based approach the bytes of the raw traffic data are converted into a gray-value images of size 28 $\times$ 28 pixels, i.e., 28 rows and 28 columns, where each pixel represents one byte of the traffic data by depicting its value as a gray-value between 0 (black) and 255 (white).

\begin{figure}[H]
\centering
    \begin{subfigure}[t]{\textwidth}
  	    \adjustbox{height=100pt, width=\textwidth}{\includegraphics{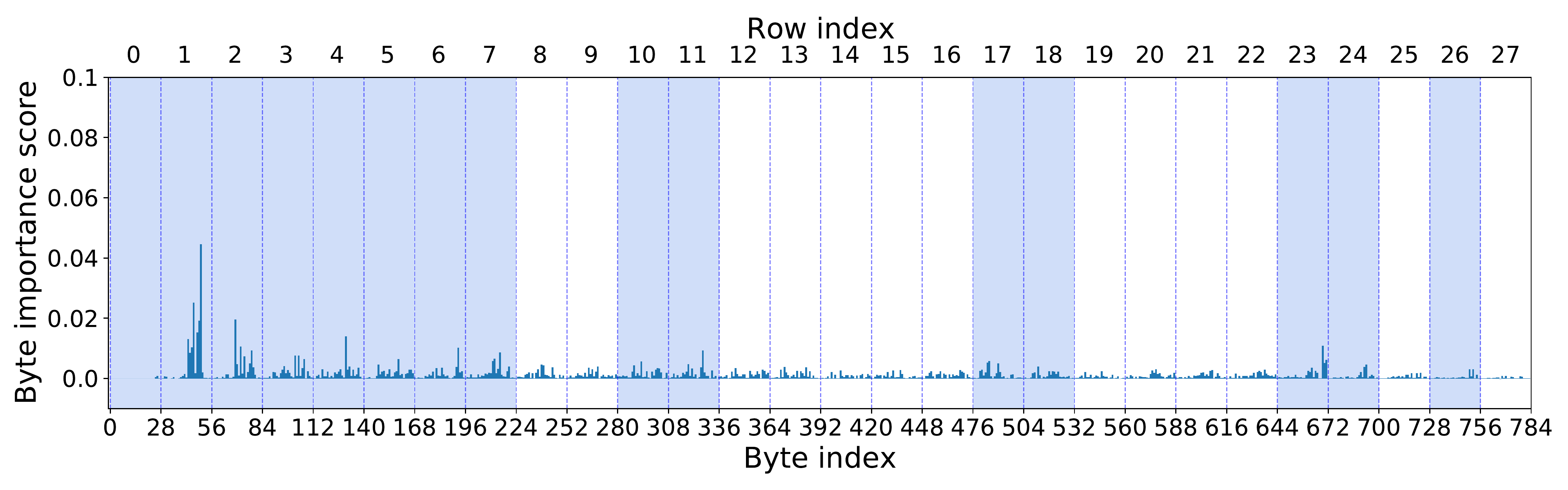}}
  	    \vspace*{-7mm}
  	    \caption{CNN}
        \label{fig:cnn_importance_barplot}
    \end{subfigure}
    \begin{subfigure}[t]{\textwidth}
  	    \adjustbox{height=100pt, width=\textwidth}{\includegraphics{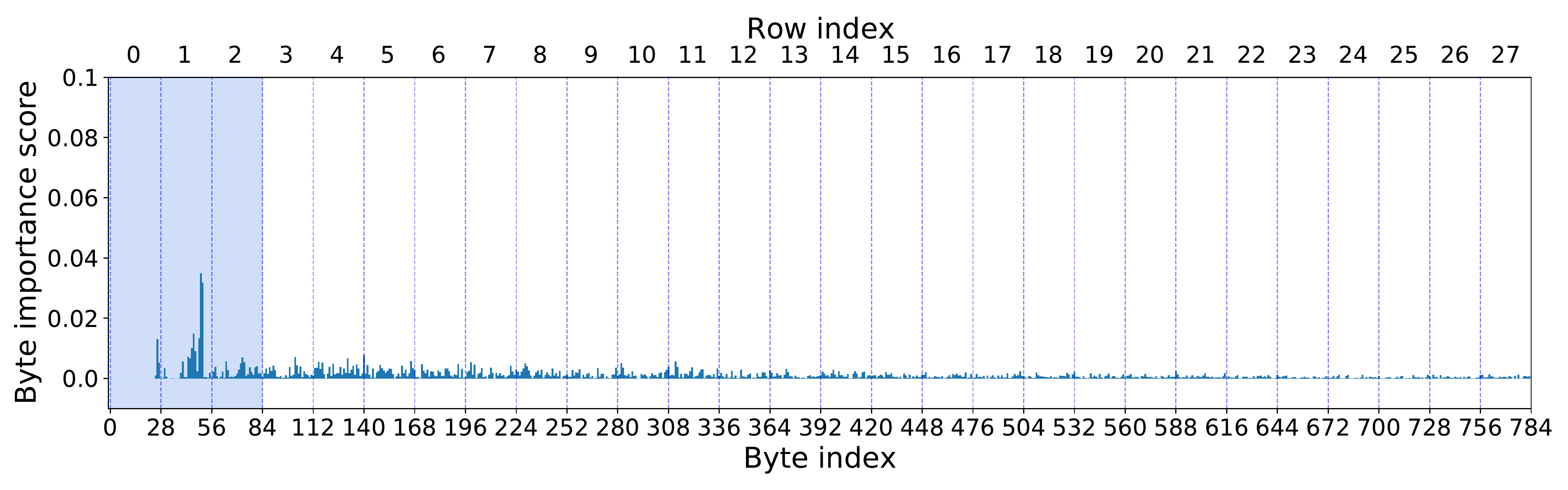}}
  	    \vspace*{-7mm}
  	    \caption{LSTM}
        \label{fig:lstm_importance_barplot}
    \end{subfigure}
    \caption{Byte importance scores for CNN and LSTM. Each grid cell represents a row in the 2D representation. Rows containing local importance peaks are highlighted. Max. value at byte index 50 for both models: 0.0446/0.0349, compared to the average value: 0.0013}
\end{figure}

Figure \ref{fig:cnn_importance_barplot} shows that the first bytes in row 1 are, on average, considered most important for device type identification with CNN. 
For individual device types, bytes in the neighbouring rows 0 and 2 also demonstrate significant importance peaks. 
This is consistent with the notion of packets located at the start of the setup protocol being of significance due to them containing header files and unencrypted data. 
Furthermore, the CNN's utilization of these bytes is coherent with Sentinel's manual fingerprint generation (cf. section \ref{sec:sentinel} which also exploits header files' data.

Figure \ref{fig:cnn_importance_barplot} additionally shows peaks for individual bytes and byte groups residing in latter rows, e.g. 3-7, 10-11, 17-18, 23-24 and 26. 
Averaging byte contributions across device types individually (in contrast to the entire Sentinel data set) shows further bytes useful for classifying a singular device type. 
This suggests that encrypted data located in subsequent rows holds information pertaining to device type specific characteristics. 
Moreover, this explains the inferior prediction performance when using manually generated fingerprint of the original Sentinel approach, which do not make use of encrypted data. 

Though LSTM's (cf. section \ref{sec:LSTM}) importance values in figure \ref{fig:lstm_importance_barplot} also display peaks in the starting rows, they constantly diminish across the input range, lacking any significant bytes/byte groups. 
This may explain LSTM's inferior performance when compared to CNN by reason of its inability to extract information from specific bytes/byte groups as well and instead focusing on a more global context, evident by the homogeneous importances seen in figure \ref{fig:lstm_importance_barplot}, especially in rows 2-7.

\section{Conclusion and Future Work}
\label{sec:conclusion}
This study contributed a methodological evaluation of two deep learning approaches, namely Convolutional Neural Networks (CNN) and Recurrent Neural Networks (RNN) in terms of the Long short-term memory (LSTM) architecture to implement the automated identification of IoT device types based on raw traffic data. 
The evaluation not only covers identification performance and runtime performance but also includes an explainability-based evaluation. 
The results are compared to an up-to-date IoT device-type identification approach using hand-crafted fingerprint features.

The identification performance shows accuracies of 97\% and 98\% for the CNN model and the LSTM model, respectively. 
Both models thereby outperform hand-crafted identification approach obtaining an accuracy of 82\%. 
The runtime performance for the IoT identification shows average runtimes of 0.2 ms and 0.67 ms for the CNN model and the LSTM model, respectively. 
Both models thereby outperform the hand-crafted identification approach by three magnitudes. 
The explainability-based evaluation shows that the LSTM model utilizes mostly the starting bytes of traffic data and acts thereby coherent with hand-crafted identification approach using that also exploits header files' data. 
The CNN model in turn also utilizes the starting bytes but additionally put emphasis on individual bytes and byte groups residing the ongoing traffic data flow. 
This suggests that encrypted data located in the subsequent data flow holds information pertaining to device type specific characteristics and therefore shows an added value for the IoT device-type identification.

Future Work should include a look at the transformer networks that are replacing more and more recurrent networks and in particular LSTM. 
Currently, transformer's networks promise to converge in training faster with  smaller training data sets compared to LSTM. 
Lastly, few-shot and one-shot learning would also serve this approach because new IoT devices could be learned only on a few samples.
\newpage
\bibliography{references}

\begin{thebibliography}{23}
\providecommand{\natexlab}[1]{#1}
\providecommand{\url}[1]{\texttt{#1}}
\expandafter\ifx\csname urlstyle\endcsname\relax
  \providecommand{\doi}[1]{doi: #1}\else
  \providecommand{\doi}{doi: \begingroup \urlstyle{rm}\Url}\fi

\bibitem[{Bengio} et~al.(2013){Bengio}, {Courville}, and
  {Vincent}]{RepresentationLearning}
Y.~{Bengio}, A.~{Courville}, and P.~{Vincent}.
\newblock Representation learning: A review and new perspectives.
\newblock \emph{IEEE Transactions on Pattern Analysis and Machine
  Intelligence}, 35\penalty0 (8):\penalty0 1798--1828, 2013.
\newblock \doi{10.1109/TPAMI.2013.50}.

\bibitem[Breiman(2001)]{Breiman01}
L.~Breiman.
\newblock Random forests.
\newblock \emph{Mach. Learn.}, 45\penalty0 (1):\penalty0 5–32, Oct. 2001.
\newblock \doi{10.1023/A:1010933404324}.

\bibitem[Cache(2006)]{cache2006fingerprinting}
J.~Cache.
\newblock Fingerprinting 802.11 implementations via statistical analysis of the
  duration field.
\newblock \emph{Uninformed. org}, 5, 2006.

\bibitem[Dainotti et~al.(2012)Dainotti, Pescape, and
  Claffy]{dainotti2012issues}
A.~Dainotti, A.~Pescape, and K.~C. Claffy.
\newblock Issues and future directions in traffic classification.
\newblock \emph{IEEE network}, 26\penalty0 (1):\penalty0 35--40, 2012.

\bibitem[Damerau(1964)]{damerau}
F.~J. Damerau.
\newblock A technique for computer detection and correction of spelling errors.
\newblock \emph{Commun. ACM}, 7\penalty0 (3):\penalty0 171–176, Mar. 1964.
\newblock ISSN 0001-0782.
\newblock \doi{10.1145/363958.363994}.
\newblock URL \url{https://doi.org/10.1145/363958.363994}.

\bibitem[Gligor et~al.(2008)Gligor, Hubaux, and
  Poovendran]{Gligor2008fingerprinting}
V.~Gligor, J.-P. Hubaux, and R.~Poovendran.
\newblock Proc. of 1st acm conference on wireless network security, wisec 2008,
  alexandria, va, usa, march 31 - april 02, 2008.
\newblock \emph{Association for Computing Machinery}, 01 2008.

\bibitem[Hochreiter and Schmidhuber(1997)]{hochreiter1997}
S.~Hochreiter and J.~Schmidhuber.
\newblock Long short-term memory.
\newblock \emph{Neural computation}, 9\penalty0 (8):\penalty0 1735--1780, 1997.

\bibitem[Kohno et~al.(2005)Kohno, Broido, and KC]{specificFingerprinting2005}
T.~Kohno, A.~Broido, and C.~KC.
\newblock Remote physical device fingerprinting.
\newblock In \emph{Remote physical device fingerprinting}, volume~2, pages
  211-- 225, 06 2005.
\newblock ISBN 0-7695-2339-0.
\newblock \doi{10.1109/SP.2005.18}.

\bibitem[Koukis et~al.(2006)Koukis, Antonatos, Antoniades, Markatos, and
  Trimintzios]{Koukis}
D.~Koukis, S.~Antonatos, D.~Antoniades, E.~Markatos, and P.~Trimintzios.
\newblock A generic anonymization framework for network traffic.
\newblock \emph{2006 IEEE International Conference on Communications},
  5:\penalty0 2302--2309, 2006.

\bibitem[Lecun et~al.(1995)Lecun, Jackel, Bottou, Cortes, Denker, Drucker,
  Guyon, Muller, Sackinger, Simard, and Vapnik]{LeCun95}
Y.~Lecun, L.~Jackel, L.~Bottou, C.~Cortes, J.~Denker, H.~Drucker, I.~Guyon,
  U.~Muller, E.~Sackinger, P.~Simard, and V.~Vapnik.
\newblock \emph{Learning algorithms for classification: A comparison on
  handwritten digit recognition}, pages 261--276.
\newblock World Scientific, 1995.

\bibitem[{LeCun} et~al.(2015){LeCun}, {Bengio}, and {Hinton}]{LeCun}
Y.~{LeCun}, Y.~{Bengio}, and G.~{Hinton}.
\newblock Deep learning.
\newblock \emph{Nature}, 521:\penalty0 436--444, 2015.

\bibitem[LeCun et~al.(2020)LeCun, Cortes, and Burges]{MNIST}
Y.~LeCun, C.~Cortes, and C.~J. Burges.
\newblock The mnist database of handwritten digits, 2020.
\newblock URL \url{http://yann.lecun.com/exdb/mnist/}.

\bibitem[Lundberg and Lee(2017)]{lundberg2017shap}
S.~Lundberg and S.-I. Lee.
\newblock A unified approach to interpreting model predictions, 2017.

\bibitem[{Marchal} et~al.(2019){Marchal}, {Miettinen}, {Nguyen}, {Sadeghi}, and
  {Asokan}]{audi2019}
S.~{Marchal}, M.~{Miettinen}, T.~D. {Nguyen}, A.~{Sadeghi}, and N.~{Asokan}.
\newblock Audi: Toward autonomous iot device-type identification using periodic
  communication.
\newblock \emph{IEEE Journal on Selected Areas in Communications}, 37\penalty0
  (6):\penalty0 1402--1412, 2019.
\newblock \doi{10.1109/JSAC.2019.2904364}.

\bibitem[{Miettinen} et~al.(2017){Miettinen}, {Marchal}, {Hafeez}, {Asokan},
  {Sadeghi}, and {Tarkoma}]{sentinel2017}
M.~{Miettinen}, S.~{Marchal}, I.~{Hafeez}, N.~{Asokan}, A.~{Sadeghi}, and
  S.~{Tarkoma}.
\newblock Iot sentinel: Automated device-type identification for security
  enforcement in iot.
\newblock In \emph{2017 IEEE 37th International Conference on Distributed
  Computing Systems (ICDCS)}, pages 2177--2184, June 2017.
\newblock \doi{10.1109/ICDCS.2017.283}.

\bibitem[{MITRE Coop.}(2020)]{CVE}
{MITRE Coop.}
\newblock Common vulnerabilities and exposures, 2020.
\newblock URL \url{https://cve.mitre.org}.

\bibitem[Nakajima et~al.(2019)Nakajima, Watanabe, Shioji, Akiyama, and
  Woo]{Nakajima}
A.~Nakajima, T.~Watanabe, E.~Shioji, M.~Akiyama, and M.~Woo.
\newblock A pilot study on consumer iot device vulnerability disclosure and
  patch release in japan and the united states.
\newblock In \emph{A Pilot Study on Consumer IoT Device Vulnerability
  Disclosure and Patch Release in Japan and the United States}, pages 485--492,
  07 2019.
\newblock \doi{10.1145/3321705.3329849}.

\bibitem[Statista(2019)]{statista}
Statista.
\newblock Internet of things (iot) connected devices installed base worldwide
  from 2015 to 2025.
\newblock 2019.
\newblock URL
  \url{https://www.statista.com/statistics/471264/iot-number-of-connected-devices-worldwide}.
\newblock Accessed: 2019-05-12.

\bibitem[Strobl et~al.(2008)Strobl, Boulesteix, Kneib, Augustin, and
  Zeileis]{strobl}
C.~Strobl, A.-L. Boulesteix, T.~Kneib, T.~Augustin, and A.~Zeileis.
\newblock Conditional variable importance for random forests.
\newblock \emph{BMC Bioinformatics}, 9\penalty0 (1):\penalty0 307, 2008.
\newblock \doi{10.1186/1471-2105-9-307}.

\bibitem[Sutskever et~al.(2013)Sutskever, Martens, Dahl, and
  Hinton]{pmlr-v28-sutskever13}
I.~Sutskever, J.~Martens, G.~Dahl, and G.~Hinton.
\newblock On the importance of initialization and momentum in deep learning.
\newblock In S.~Dasgupta and D.~McAllester, editors, \emph{Proceedings of the
  30th International Conference on Machine Learning}, volume~28 of
  \emph{Proceedings of Machine Learning Research}, pages 1139--1147, Atlanta,
  Georgia, USA, 17--19 Jun 2013. PMLR.
\newblock URL \url{http://proceedings.mlr.press/v28/sutskever13.html}.

\bibitem[Wang et~al.(2017)Wang, Zhu, {Xuewen Zeng}, {Xiaozhou Ye}, and {Yiqiang
  Sheng}]{WeiWang2017}
W.~Wang, M.~Zhu, {Xuewen Zeng}, {Xiaozhou Ye}, and {Yiqiang Sheng}.
\newblock Malware traffic classification using convolutional neural network for
  representation learning.
\newblock In \emph{2017 International Conference on Information Networking
  (ICOIN)}, pages 712--717, 2017.

\bibitem[{Werbos}(1990)]{BPTT}
P.~J. {Werbos}.
\newblock Backpropagation through time: what it does and how to do it.
\newblock \emph{Proceedings of the IEEE}, 78\penalty0 (10):\penalty0
  1550--1560, 1990.

\bibitem[Williams and Zipser(1995)]{Williams95gradient-basedlearning}
R.~J. Williams and D.~Zipser.
\newblock Gradient-based learning algorithms for recurrent networks and their
  computational complexity, 1995.

\end{thebibliography}

\end{document}